# A Focused Crawler Combinatory Link and Content Model Based on T-Graph Principles

Ali Seyfi

[1]Department of Computer Science
School of Engineering and Applied Sciences
The George Washington University
Washington, DC, USA.

[2]Centre of Software Technology and Management (SOFTAM)
School of Computer Science
Faculty of Information Science and Technology
Universiti Kebangsaan Malaysia (UKM)
Kuala Lumpur, Malaysia

seyfi@gwu.edu

**Abstract**— The two significant tasks of a focused Web crawler are finding relevant topic-specific documents on the Web and analytically prioritizing them for later effective and reliable download. For the first task, we propose a sophisticated custom algorithm to fetch and analyze the most effective HTML structural elements of the page as well as the topical boundary and anchor text of each unvisited link, based on which the topical focus of an unvisited page can be predicted and elicited with a high accuracy. Thus, our novel method uniquely combines both link-based and content-based approaches. For the second task, we propose a scoring function of the relevant URLs through the use of T-Graph (Treasure Graph) to assist in prioritizing the unvisited links that will later be put into the fetching queue. Our Web search system is called the Treasure-Crawler. This research paper embodies the architectural design of the Treasure-Crawler system which satisfies the principle requirements of a focused Web crawler, and asserts the correctness of the system structure including all its modules through illustrations and by the test results.

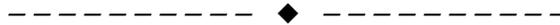

## 1 INTRODUCTION

The openly indexable World Wide Web has empirically passed twenty billion ($2\times10^{10}$) Web pages [1] and hitherto, growth does not appear to level off. Considering the exponentially increasing amount of dynamic content on the Web, such as news, schedules, social networks and individual data, it is evidently asserted that relying on search engines is inevitable for the people to approach their desired information. Whereof these concerns make searching the Web a profound task, experts apply a lot of machine learning methods to accomplish several phases of this job such as ranking retrieved Web pages based on their estimated relevance to the user query. The main goal is to make up the best weighting algorithm that could represent queries and pages as in a vector space. This way, the closeness in such a space would convey the semantic relevance.

A Web crawler systematically collects information about documents from the Web to create an index of the data it is searching and it maintains an updated index through subsequent crawls. As an automatic indexer, the crawler operates in the context of listing the documents relevant to a subject or topic which one would expect in a typical user search query. Traditional general purpose Web crawlers are not easily scalable since the Web is not under the control of one operator or proprietary. They also may not be set to target specific topics for accurate indexing, and lag behind in time and updates to manage updated indexes/indices of the whole Web to stay current because of the distribution, subject and volume involved. To overcome these shortcomings, focused crawlers are intended to rely on the linked structure of the Web in order to identify and harvest topically relevant pages to increase their performance in terms of accuracy, currency and speed. A significant benefit in using focused crawlers is the possibility of decentralizing the resources and storage indexes.

There are two major open problems in focused crawling: the first problem is the prediction of the topical focus of an unvisited page before actually attempting to download the content of the page. As one of its fundamental tasks, the crawler should use specific algorithms in order to make this prediction with the highest possible accuracy. Typically, the focused crawlers download the whole content of the page, analyze it and make a decision on whether it is related to their topic of interest. Alternatively, some researches show that the topical focus of a page can be predicted by analyzing the anchor text of its link in the parent page. Between these two extremes, in our research we take into account several HTML structural elements of the parent page, in addition to the anchor text. This will help improve the accuracy of the topical detection of the unvisited link. The second problem is prioritizing the links for later downloads. A proper priority score should be assigned to all the extracted URLs from a Web page. These URLs along with their scores will then be put in a queue to be downloaded later. This task of prioritization is very important in that some irrelevant pages should be visited and passed off on the way to reach another populated area of relevant pages. For prioritization, we employ a novel tree-like data structure called T-Graph (Treasure Graph) [2]. The traversal method in the construction of T-Graph can be both top-down or bottom-up. Also, in our proposed system which is called the Treasure-Crawler, no resources from any other search engines are required, provided the T-Graph is set to be constructed top-down.

This study is conducted to determine a novel crawler's effectiveness and viability in crawling to fetch topic specific documents. The system is designed and implemented in a way that all the components/modules have the minimum dependency on each other; hence each can be changed or replaced while requiring the least manipulation of other parts except for some interfaces. The document standards such as HTML and XML are respected for this system. Also HTTP and HTTPS communication protocol standards from IETF and W3C are considered, as well as the robot exclusion protocol known as "robots.txt".

### 1.1 The Methodology

One of the main objectives of this research has been to enhance the accuracy of Web page classification, which is made possible by defining a flexible interpretation of the surrounding text, in addition to specific HTML data. The Dewey Decimal Classification (DDC) system is applied as a basis to classify the text into appropriate topic/subject boundaries. The other significant objective of this research has been to reach target documents in the shortest possible time. This is accomplished by reaching the T-Graph most matching node(s) with the document, and then calculating the distance of such nodes to the target level based on the number of documents to download [2]. As a result, due to a better determination of the topic boundary and significant decrease of the volume of downloaded documents in text format, this strategy helps the crawler update its indexes more pragmatically, accurately and rapidly. Based on these assumptions, we designed a new algorithm and built a prototype to exercise, test and validate it against the functional and non-functional requirements. The summary results are only presented in this paper, while the actual detailed evaluation results are reported in a follow-up paper with the title of "Evaluation of Link and Content Based Focused Treasure-Crawler".

In the rest of the current paper, we consecutively review some major Web crawlers and their reported experimental results, detail the requirements and evaluation criteria of a focused Web crawler, present the architectural design of the Treasure-Crawler, elaborate the employed methods and algorithms, describe the outcome of the carried-out tests and validations, and conclude with a summarized list of results and conclusions.

## 2 BACKGROUND

Topical and focused crawling are two slightly different concepts, first introduced by Menczer [3] and by Chakrabarti [4] respectively. A focused crawler selects and seeks out pages that are relevant to a pre-defined topic (or a set of topics). It systematically analyzes its crawl frontiers and tries to detect the pages that are most likely to be on the designated topic(s) of the crawl, while it tries to avoid off-topic Web regions. As a result, this process brings considerable savings in different resources, such as network, computing, and storage, hence, the crawl becomes more up-to-date. Usually, the topics of interest are optionally defined by keywords, categorized/classified standard lexicon entries, or by a set of exemplary documents. The major challenge of a focused Web crawler is the capability of predicting the relevance of a given page before actually crawling it. Achieving this goal requires particular intelligence for the crawler. Focused Web crawlers use this kind of intelligence to avoid irrelevant regions of the Web in order to make the task manageable. Additionally, a focused Web crawler should also pay attention to the ability to discover relevant regions which are separated by groups of irrelevant Web regions in order to achieve desirable Web coverage. A well-designed focused Web crawler should be able to stay in pre-defined topics as long as possible, while covering the Web as much as possible.

To index the Web, the first generation of crawlers relied on basic graph traversal algorithms, namely the breadth-first or depth-first. An initial set of URLs are used as a seed set, and the algorithm recursively follows hyperlinks down to other documents. Document content is of less importance since the ultimate goal is to cover the whole Web. A focused crawler, on the other hand, efficiently seeks out documents about a specific topic and guides the search based on both the content and link structure of the Web. The main strategy is to associate a score with each unvisited link within the pages it has downloaded [5, 6].

**Best-first** is basically an optimization to the Breadth-first algorithm. When the unvisited links are extracted, an estimator tries to prioritize them. After being associated with a priority score, these links are inserted into a priority queue. These links are then fetched from the queue according to their assigned priority.

**Panorama (1996)** is one of the first systems that established a digital library by using the Web and made CiteSeer, the most popular Web-based scientific literature digital library and search engine, primarily focusing on information technology and computer science. Panorama traverses the Web for relevant documents in PDF and Postscript formats in the computer science field. To construct its topic of interest, Panorama submits relevant papers' main titles and the titles of references within the papers as separate exact phrase queries to Google Web APIs [7]. Thus, the returned URLs form a positive example set. Examples from unrelated papers form a negative example set. Both of the two sets are used to train a Naïve Bayes classifier, which is used to guide the crawling process [8].

**Shark Search Crawler (1998)** tries to enter the areas where a higher population of relevant pages is observed, but stops searching in the areas with no or very little number of relevant pages.

**InfoSpiders (1999)** constitutes a population of adaptive intelligent agents. These agents use neural networks algorithms and are very advanced in distinguishing fruitful links to follow.

**Context Focused Crawler (2000)** builds upon constructing a multi-level tree of sample documents called the context graph. This algorithm, first proposed by Diligenti et al. [9], has been an active research area in the past decade.

Fig. 1 shows a context graph. For each target document (e.g. P) one such graph is constructed. As the target documents are supplied to the system in Level 0, another search engine is utilized to find the high rank pages that contain a link to the current target document (e.g. A and B) and they are put in Layer 1. This process is then recursively repeated for each parent page until the graph reaches a desired number of levels. By convention, there is no connection between the nodes in a common layer. Also, if two (or more) documents in layer i (e.g. A and B) have a common parent in layer i+1 (e.g. C), the parent should appear two (or more) times in layer i+1. Note that Layer 1 Web pages are one step away from desirable documents and Layer 2 Web pages are exactly two steps away from desirable documents and so on. When the context graphs are constructed they are merged to a single graph called the merged context graph. If there are n layers in the merged context graph, n classifiers are trained to assign Web pages into its different layers. The assigned level to each page defines its priority, which is the closeness of the page to the desired topic. As a result, the context graph can learn topics that are indirectly connected to a pre-defined topic and thus increases recall.

As expected, in practice the arrangement of the nodes in the layers of the context graph reflect any hierarchical content structure. Highly related content typically appears near the center of the graph, while the outer layers contain more general pages. The idea of using the context of a given topic to guide the crawling process could significantly increase both precision and recall.

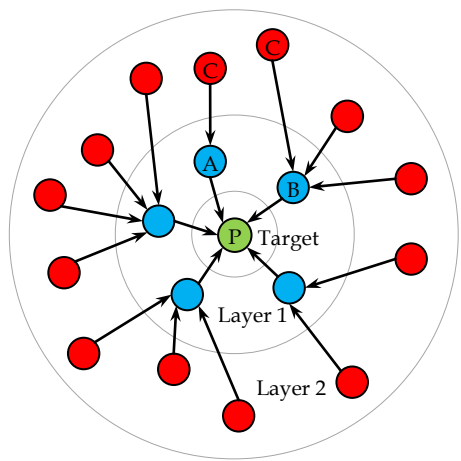

Fig. 1. A context graph

**Tunneling (2001)** is the phenomenon where a crawler reaches some relevant regions (or pages) while traversing a path which does not solely consist of relevant pages [10, 11]. The major task of focused Web crawlers is to unveil as many bridges among relevant regions as possible. Martin Ester et al. implemented their tunneling technique based on taxonomies. During the crawling process, the crawler keeps an eye on the precision. If the precision goes down much faster than expected, it is necessary to broaden the focus of the crawl. For example, if a focused Web crawler is currently working on the topic "basketball" and cannot find more relevant pages, it can generalize the topic to "sport" and expect to find more bridges leading to other relevant regions. If the precision gets better than a pre-defined value after the generalization, the topic is specialized again. Taxonomy based tunneling seems an ideal and intuitive approach to increase recall.

**Meta Search Crawler (2004)** is designed to build domain-specific Web collections for scientific digital libraries [12]. The author addressed the drawbacks of the traditional focused Web crawlers caused by local search algorithms in the digital library construction. The meta-search crawler follows a novel approach; generating queries from a domain-specific lexicon and combining the entries on the first page of the results from multiple search engines. This way, the crawler will cover diverse Web communities of documents which are presumed to have very little overlap. This approach is very effective in terms of the accuracy in remaining on the topic while showing a high Web coverage. The experimental results of the meta-search crawler show the maximum precision value of 50.23% which is almost equal to the precision value of our Treasure-Crawler. However, it is controversial if using and relying on other search engines is effective or not.

**LSCrawler (2006)** considers better recall as its goal and uses link semantics to guide the crawling process [13]. The LSCrawler takes into account the semantic of anchor text of the link in order to specify the relevancy of the unvisited page. Since anchor texts are often too short to provide adequate information to arrange a reasonable crawling sequence, content texts and URL patterns are optionally considered as well. However, in their paper, Yuvarani et al. do not propose a clear description of the surrounding text. LSCrawler is basically a semantic crawler as it utilizes the corresponding ontology of its topical focus to make assessments on a URL. Although the structure of this algorithm has useful commons with our proposed algorithm, relying on its experimental results may not be insightful because of its semantic basis. They report a recall value of ~0.6 which is a good result although the number of the crawled pages is not indicated and an unlimited increase of the recall value is observed, which should normally approach a constant level.

**Hybrid Focused Crawler (2006)** uses two features of documents; link structure and content similarity of each page to the main topic to crawl the Web [14].

Initially the topic of interest is set for the crawler to launch the crawling using a single seed URL. All Web pages are mapped into one of three groups: Seed Pages, Candidate Crawled Pages and Uncrawled Pages. Seed Pages are repeatedly selected from Candidate Crawled Pages using a rank score. Only hyperlinks derived from Seed Pages are crawled. All crawled pages are first stored in the Candidate Crawled Pages table and their rank scores are calculated after each session. The one with the highest mark will be moved to the Seed Pages table. The rank score is calculated using a combination of the "link structure" and "content similarity" as below:

$$\text{Rank}(p) = (\text{links\_to\_seed}(p) + \text{links\_from\_seed}(p)) \times (0.1 + \text{content\_similarity}(p)) \quad (1)$$

where p is a page from Candidate Crawled Pages; links_to_seed and links_from_seed are respectively the number of links from the page p to Seed Pages and the number of links from Seed Pages to it; content_similarity is a real number between 0 and 1 indicating the similarity degree of a page to a topic. From the above formula, we can observe that this algorithm emphasizes the link structure over content similarity and it is essentially a compromise of the PageRank algorithm [15]. Using Seed Pages to mimic important pages in the PageRank algorithm is a promising way to remove the requirement of a large on-hand Web link structure database but it is also a potential limitation where the accuracy of this algorithm largely depends on the precision of Seed Pages. The content_similarity is calculated using vocabulary vectors and simple keyword matching. They deemed that its classification accuracy cannot be as good as SVM classifiers or Naïve Bayes classifiers. In addition, using only one seed URL to start the crawling process is an obvious flaw.

The Web crawler proposed by Jamali et al. [14] demonstrates a high harvest ratio in the retrieval of on-topic pages, although it follows a rather different methodology than our proposed system. Their system is intended to first download and then determine the relevancy of a Web page. This approach, however, is presumed to have a high overload on the performance of the system as the Web is rapidly increasing in size. In their experiment, Jamali et al. reported a harvest ratio of 67.2% ~ 85.5%, which is a significant outcome after crawling 25000 Web pages.

**Relevancy Context Graph (2006)** is designed based on the context graph algorithm [16]. There are two major problems with the context graph algorithm. The strict link distance definition of this algorithm brings too much classification burden on the system. Also, this homogenous link requirement lowers the classification accuracy because it does not tally with the real topology of all the topics on the Web. To overcome these problems, Hsu and Wu [16] proposed the relevancy context graph structure and reached an approximate value of ~0.22 for the precision in the category of "mountain". The problem with this approach is that the selected topics may be at different levels of specificity.

**HAWK (2008)** is simply a combination of some well-known content-based and link-based crawling approaches, and results in a decreasing value of harvest ratio (0.6) after crawling 5000 pages [17]. This val-ue is quite acceptable; however, its decreasing rate is daunting because it is not clear what constant value it will gradually reach.

**Modified NB (2010)** implements a search system, where the classifier utilizes reinforcement learning in order to train itself against topic taxonomy. It also stems and assigns proper weight to the textual data as well as to the URLs and anchor texts [18]. For this weight assignment, it uses a modified form of TF-IDF. For a better classification, an adopted form of Naïve Bayes classifier is employed. The experimental results of this system, in comparison with our proposed system, show a better harvest ratio.

**Intelligent Focused Crawler (2011)** employs a modified form of the Naïve Bayes classifier and results in a significant harvest ratio. For this crawler, Taylan et al. investigated several link scoring methods, namely an updatable form and a threshold-restricted form of the Naïve Bayes classifier [19]. Their fruitful results demonstrate that the Naïve Bayes with threshold approach gives a higher harvest ratio among other algorithms. The two modifications they have made to the Naïve Bayes classification method improve the crawling performance in terms of the harvest ratio (around 80%), which in comparison with our system is much higher and more effective. The problem with their experiment is that they have only crawled 1000 Web pages which is not a satisfactory number to represent the whole Web.

**OntoCrawler (2011)** follows a fuzzy approach in weighting the concepts and page contents, while relying on ontologies in the semantic Web [20]. It demonstrates an increasing precision value in a crawl of 1000 pages, unlike other previously introduced algorithms in focused crawling. To demonstrate its improved performance, the precision rate of 90% of the Onto-Crawler was compared to the resulted precision of the Best-First and Breadth-First crawlers on the topic of football. Although our proposed system has only a precision of ~50%, it is not boosted with fuzzy logic and semantic Web capabilities as in the OntoCrawler.

## 3. REQUIREMENTS AND EVALUATION CRITERIA

In this section we define the requirements of a focused crawler through the use of scenarios describing how the system is used in practice. These are helpful in requirements elicitation, elaboration and validation. To describe the scenarios of the proposed system, the Viewpoint-Oriented method (VORD) [21] is used with the following steps:
1. System admin provides the data structure (tree or graph) with a sample data set of interlinked Web pages, and also provides the crawler with seed URLs.
2. Data structure automatically downloads the required information from the Web to build and update itself. This step may be taken repeatedly in the crawling process.
3. Crawler fetches the requested Web pages from the Web.
4. Crawler hands the downloaded Web page to the main relevance calculator of the search engine.
5. Relevance calculator uses its internal algorithms and techniques along with the output from the data structure to make specific decisions on the content of the page and its unvisited URLs.
6. Web page content and all the related data are stored in the search engine repository.
7. Extracted URLs from the Web page are supplied to the crawler.

The complete, detailed and unambiguous specifications and descriptions of what the system is supposed to do form the system requirements.

### 3.1 Functional Requirements

The list of functional requirements consists of the services that the system should provide, the system's reactions to particular inputs and the system's behavior in particular situations:

**System initialization:** The system shall start from a stable point provided that all the initial inputs are supplied to the system. The system shall then receive the required inputs such as the seed URLs, documents, values and thresholds. The queues and all data structures shall be emptied, initialized and fed with the required initial data that are provided by the system administrator. All the system's hardware should be checked, including all types of server(s) and the network connection(s) along with their performance. The starting pages (seed URLs) shall be downloaded correctly and be ready for usage.

**Relevance calculation:** The system shall be able to calculate the relevance of a Web page based on its textual data and assign a topic code as in a specified classification system. The URLs in a Web page and their topical boundaries should be detected and extracted. All the text data should be stemmed before any calculation. The topical focus of the stemmed textual data should be inspected by using a comparative algorithm.

**Priority calculation:** The system shall use a specific structure to assign a priority score to each unvisited link present in the downloaded Web page. The graph structure should be initialized by receiving a sample data set of interlinked documents, or only the target documents if it is designed to build itself automatically. The graph structure should search for and download the relevant pages in order to fill the nodes. This process should continue correctly until the desired number of levels and nodes is gained. The HTML data of the downloaded Web page should be extracted in order to be compared with the data in the graph nodes. This comparison needs a specific text similarity measurement algorithm. The overall priority of the unvisited link should be calculated by using a specific formula. This priority score should be then assigned to the URL.

**Data storage and indexing for later search:** If the downloaded Web page is recognized to be relevant to the specialized topic of the system, it shall be stored in the search engine repository. The contents of the repository are then fetched and presented in response to specific user queries. Relevant Web pages shall be stored in the search engine repository along with all their related information that is provided or calculated by the system.

### 3.2 Non-Functional Requirements

Non-functional requirements are the constraints and limitations on the services or functions offered by the system, such as I/O device capabilities and system representations. They also define the system's properties such as reliability, response time and storage requirements. It must be stated that non-functional requirements may be more critical than functional requirements. If they are not met, the system may be useless as for safety measures such as trust, protection, audit trails, security, privacy, digital forensics and identity record management [22]. The non-functional requirements of a Web crawler are as follows:

**Flexibility:** The system should be usable in various scenarios with the least possible modifications.

**Low cost:** The system should run on low-cost hardware and the number of servers should be minimized, provided that the overall performance is not affected.

**High performance:** Basically the system should be able to process several hundred pages per second and several millions per run. Therefore, the disk access and usage is vital, especially when the data structures such as queues and graphs overfill the memory. This situation occurs when millions of pages are downloaded.

**Robustness:** Dealing with the Web as the largest existing collection of data and information, the crawler system interacts with millions of servers and pages. Thus, it has to tolerate corrupted HTML data, unexpected server configurations and behavior, and many other strange issues. Since each crawl consumes weeks or even months of time, the system should be able to handle network crashes and interruptions with losing the least possible amount of data.

**Standardization:** Following the standard conventions is highly important for a crawler. The crawler should respect the standards such as "robots.txt" which is a robot exclusion protocol [23] and "sitemaps" which is a robot inclusion protocol [24]. In addition, supplying the crawler with a contact URL and supervising the crawl are very important.

**Speed control:** The system should match the crawl speed with the load of its main routers. A sliding window can limit the crawl speed at peak usage hours and increase the speed during the lower usage hours.

**Manageability:** The system should have a proper interface to show the monitoring information, such as crawler speed, size of main data sets, and some statistics about pages.

**Reconfigurability:** The administrator should be able to adjust the speed, add and remove components, shut down the system, force a checkpoint, or add hosts and domains to a "blacklist" of places that the crawler should avoid. Moreover, after a crash or shutdown, the software of the system may be modified to fix problems, and we may want to continue the crawl using a different machine configuration.

**Modularity:** The system should be designed and implemented modularly. This will help easy and rapid accommodation of changes in the configuration of the system.

**Reliability:** For each crawl of a specific number of pages, the output of the system should be reliable to be used for making further changes in the system. Also it should be guaranteed that the downloaded pages are identified and prioritized in the best possible way.

**Software and hardware independency:** The crawler needs to access pages that are situated on various servers with various platforms. Therefore, it should be flexible and act independent of the hardware and software it is encountering.

### 3.3 Evaluation Criteria

Focused crawlers share many commons with the traditional information retrieval systems and are therefore evaluated regarding the criteria that apply to all of the novel and traditional IR systems. Moreover, some of these criteria are set to be measured within the context of software architecture as well as the Web and its attributes. Although only some of these criteria are measured for academic researches, all of them should be satisfied to put a Web crawler into work. It may be possible to introduce new metrics per use, especially when time constraints are vital. In order to collect data for these evaluations, our implementation of the Treasure-Crawler records all its activities as the system is run and in a timely manner. For this purpose, all HTTP responses, detailed calculations and even temporary results are stored in a database to perform further off-line analysis. A follow up paper will elaborate these results, analysis and evaluations.

**Harvest ratio:** This is the rate at which relevant pages are detected and high-prioritized, and irrele-vant pages are efficiently discarded or low-prioritized. This value should reach a threshold; otherwise an exhaustive crawler may be preferably used.

**Recall:** This is the proportion of relevant retrieved Web pages to all the relevant pages on the Web. To approximate the system recall, we presumed our crawled pages as the whole collection and only took the pages that were topically known within the DMOZ dataset [25].

**Precision:** This is the proportion of relevant retrieved Web pages to all the retrieved Web pages. This metric should also be high.

**Performance:** This is a general term to use but what is important is that the system should have good performance in terms of speed and other IR accuracy metrics.

**Scalability:** In any scale, from a small dataset to the huge Internet, the system should be adjustable and scalable. Our prototype is implemented to scale the Web as well as any provided dataset of documents and URLs, such as the DMOZ, on which it has been run and tested. However, it is not evaluated against this criterion because the results of such examination will not be substantial unless actual parallel Web servers and high bandwidth Internet connection are available.

**Modularity:** The system should be implemented by respecting as many standards as possible, to provide a flexible and plain basis for further developments on other interfaces and mechanisms. In the case of replacement of any components of the system, the other parts should require the least change. Although this criterion is not evaluated in detail, the prototype system of this study is specified, designed and implemented to follow software engineering modularity philosophy whereby one module can be easily replaced with another new one without affecting any other modules of the system. Since the internal working functions of the modules are self-contained, any required changes could be only at the interface between modules if a new or unwanted parameter exchange demands it. As we are following the object oriented approach, the classes and their objects are designed and interconnected in such a way to flexibly accommodate changes and replacements.

**Robustness:** The system has to tolerate corrupted HTML data, unexpected server configurations and behaviors and any other strange issues and errors. Our system detects and records these situations and makes specific decisions or provides error messages for the system administration function.

Our proposed system is implemented as a prototype and tested on a typical machine with a permanent high speed Internet connection. The network, hardware and software conditions of such experimental prototype limited our evaluations to focus on the information retrieval concepts, hence, only those criteria that represent the accuracy of the retrieval will be measured and presented. It must be stated that the prototype is designed and implemented in such a way to satisfy those criteria that relate to its software engineering fundaments (e.g. scalability, modularity, and robustness).

## 4. FRAMEWORK AND ARCHITECTURE

The conceptual framework architecture of the Treasure-Crawler-based search engine is illustrated in Fig. 2, followed by a description of the modules. To target the design optimality, the Treasure-Crawler architecture and prototype was tested and evaluated against a set of criteria, and the results are discussed in a follow-up paper. However, a summary of the results is given in Section 6.

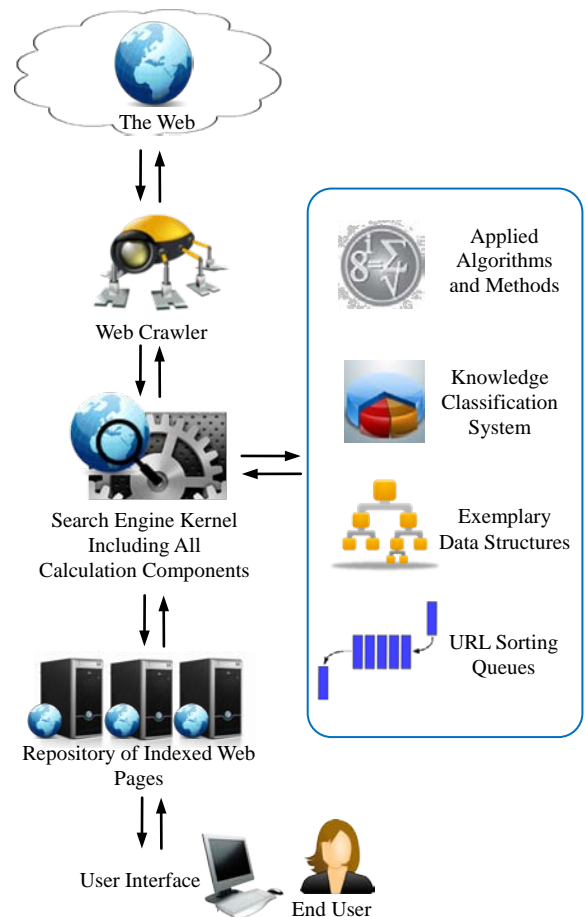

Fig. 2. Conceptual framework of the Treasure-Crawler-based search engine

*Web Crawler* is a software agent that receives a URL from the URLs queue (supervised by the kernel) and downloads its corresponding Web page. The page is then parsed and analyzed by the kernel to be ready for the next stages. If the crawler receives any other HTTP response instead of the page content, it tries to store the URL along with its specific message.

*Search Engine Kernel* is the main computational component of the search engine. All the decision makings and necessary calculations are performed in the kernel. It controls all other components while utilizing their data. In focused crawling, the relevancy of the pages and priority assignment to each URL are the main duties of the kernel. In our proposed architecture, the kernel predicts the topical focus of an unvisited link through analyzing its anchor text and the content of the parent page. Thus, the system is set to combine link-based and content-based analysis.

*Applied Algorithms and Methods* are variously utilized to achieve the ultimate goal of the system, which is to effectively index and search the Web. In our system, these custom methods are designed to be simple while efficient. Alternatively, one may use other predefined algorithms such as specific classifiers and text similarity measurements.

*Knowledge Classification System* is used as a reference to classify each Web page or unvisited URL into its appropriate sub-category of human knowledge. In the context of information retrieval and library science there are several classifications systems that are widely used. For our system we chose Dewey Decimal system according to its wide range of knowledge coverage.

*Exemplary Data Structure* is a reference hierarchy based on which the priority of the URLs can be calculated. This reference hierarchy is basically a data structure of some sample pages. Similar to our utilized T-Graph, one can design and use a custom structure for this purpose.

*URL Sorting Queues* are key data structures of any search engine, wherein URLs are lined to be utilized in the intended manner. They are mostly organized as priority queues; however first-in-first-out structure is also widely used.

*Repository* stores an indexed version of the Web. Although this repository cannot keep the live version of the Web, it must be indexed for fast retrieval. Also the repository should stay as up-to-date as possible by use of an intelligent and fast module.

*User Interface* is the only user-side component of a search engine. This interface should be graphically designed in such a way to receive the search query easily and present the search results quickly and clearly.

Given the above framework, the architectural design of the Treasure-Crawler Web search system is shown in Fig. 3, followed by the descriptions of the processes in a sequential order.

1. The crawler component dequeues an element from the fetcher queue, which is a priority queue. Initially

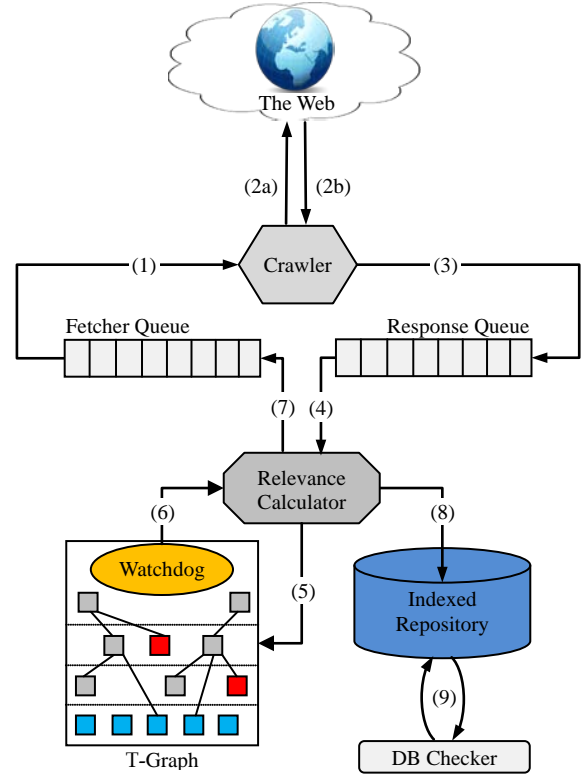

Fig. 3. Architectural design of the proposed Treasure Crawler Web search system

the seed URLs are inserted into the queue with the highest priority score. Afterwards, the items are dequeued on a highest-priority-first basis.

2. (a) The crawler component locates the Web documents pointed by the URL taken from the fetcher queue and (b) attempts to download the actual HTML data of the page, or otherwise, the server's HTTP response.

3. For each document downloaded, the crawler places the response in the response queue. The response queue contains the documents or HTTP response, in case the page cannot be downloaded due to temporary unavailability or non-freshness of the link.

4. The document is then handed to the relevance calculator which processes the documents and analyzes whether the document belongs to the specialized topic or not.

5. If considered as on-topic, particular elements of the page are then sent to the T-Graph to carry out specific comparisons and calculations. The T-Graph data is used to determine the importance of the unvisited links within the page.

6. T-Graph associates a priority score to each unvisited link. Even the off-topic URLs are assigned a lowest value as the priority. This helps the crawler to harvest unlinked on-topic regions on the Web, which are connected through off-topic pages. There is an observing component called the watchdog on the T-Graph. It periodically updates the T-Graph in order to incorporate its experiences.

7. After completing all analyzes, the relevance calculator inserts the URLs and their assigned priority scores to the fetcher queue. The priority score of the fetcher queue items is cyclically incremented to prevent starvation.

8. HTML data of the analyzed Web page is fully stored in the repository along with all the measurements such as the priority scores given to the links.

9. A database checker component constantly runs specific checkups on the repository and updates its indexes with the ultimate goal of keeping the repository up to date.

Note that depending on software engineering object oriented concepts and Web application operational constraints and regulations [21], our system framework and architecture respects all the standards both related to modular programming and W3C regulations [26].

## 5. PRACTICAL ANALYSIS

There are two significant tasks for a focused Web crawler; detecting the topical focus of an unvisited page and assigning an appropriate priority score to its URL. This is achieved by using the T-Graph structure and other custom methods described below.

There are two significant tasks for a focused Web crawler; detecting the topical focus of an unvisited page and assigning an appropriate priority score to its URL. This is achieved by using the T-Graph structure and other custom methods described below.

### 5.1 Detecting the Topical Focus

As the knowledge classification system, we exploit the Dewey decimal system [27, 28], where the human knowledge is divided into super-categories, each of which is divided into 10 categories and further divided into 10 sub-categories. After the third digit, there is a decimal point and the numbering continues to take a topic into more detail and specificity. For our prototype, we only focused on the first 3 digits and not the decimal part. However, the system is implemented in such a way that any length of the codes can be considered, provided that the complete version of the Dewey system is available. We call each code of the Dewey system a D-number.

Starting from the words of an unvisited link in the current Web page, a (conceptual) 2-D graph is plotted with the complete list of D-Numbers and their length as the dimensions. Before any textual comparison, the text is stemmed using the Porter stemming algorithm [29]. This way, the textual data is normalized and the errors are minimized.

In order to segregate the textual data of the Web page into words/phrases, an HTML document parser is used. It gives phrases if any phrase is directly present in DDC, otherwise it gives the words. It is possible for a word/phrase to have more than one D-Number. The word "clothing" is an example which belongs to several disciplines. It belongs in 155.95 for its psychological meaning; customs associated with clothing belong in 391 as part of the discipline of customs; and clothing in the sense of fashion design belongs in 746.92 as part of the discipline of the arts. Therefore, for each word/phrase, more than one point may be plotted. The plotting is stopped when the paragraph boundary is reached (called break points, which initially could constitute the starting and ending of the paragraph itself). If the unvisited link is a list item, then also the plotting is stopped after plotting all the list items. Then the plotted points are analyzed. [2]

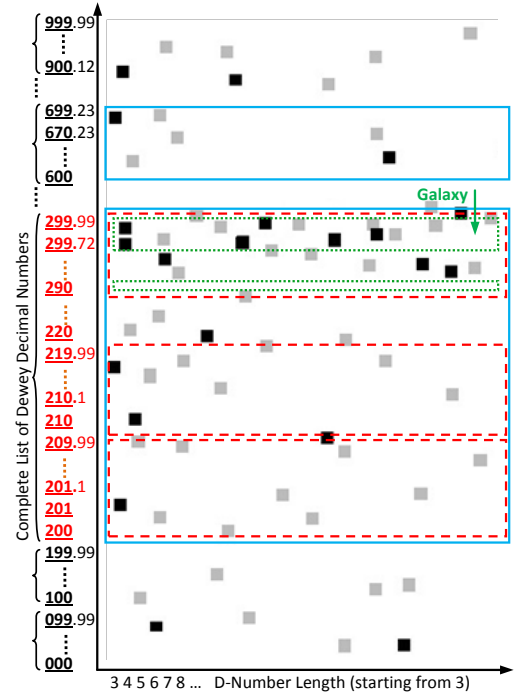

Fig. 4. Example of processing the D-Numbers and retrieving the galaxy

Fig. 4 shows an example of how a graph looks when plotted along with the explanation of the concept as follows:
• All the dots constitute the words between the break points.
• Each light colored (un-bolded black or grayish) dot shows the point corresponding to the word, which is not the anchor text. Each dark colored (bolded black) dot shows the point corresponding to the anchor text of the unvisited link.
• The target is to find the area with the highest popu-

lation of dots called the Galaxy, which will show the topical focus of the text.

Although there are many methods in the Data Mining subject field to eliminate the outliers from a sample data set [4], in our system a simpler method is used to detect the Galaxy, while taking several factors into account. This algorithm takes three characteristics of the words present in the Galaxy. First is the number of words which determines the thickness of the population. Second is the D-Number length which explains the preciseness of the topic. Third is called the anchor impact, where black colored dots have a higher impact since they belong to the words in the anchor text. These words have an effect of about 40% more than the text around the unvisited link [30].

To graphically demonstrate the process, rectangular regions are shown in Fig. 4. We define W as the weight of each rectangle. The blue (lined) rectangles show the regions where D-Numbers have the same first digit. For all of these regions (ten regions), W is calculated and the rectangle with the maximum value of W is considered. For example, the rectangle for D-Numbers starting with 2 is selected which is related to the super topic "religions" as in DDC. It is also possible to consider more than one region with the highest W. Then only the dots (D-Numbers) inside the selected blue rectangle are considered. This time, the rectangular regions (red dashed rectangles) are selected based on the second digit (ten regions) and again for all of them W is calculated. The one(s) with the highest value of W are considered. In this example, the selected region constitutes D-Numbers with 2 as the first and 9 as the second digit, which is related to "other religions" as in DDC.

Again, the dots in the selected red rectangle(s) are divided into ten smaller rectangles, according to the third digit of their D-Number (green dotted rectangles) and W is calculated for each of them. Since in this example the process has to continue until three digits, the green rectangle(s) with the maximum value of W is (are) considered as the Galaxy and show the topical focus of the text, which in this example is "Religions not provided for elsewhere" according to DDC. Parameter W is calculated as follows:

$$W = n \times \sum_{i=1}^{n} \{length(d_i) \times anchor\_impact(d_i)\} \quad (2)$$

where:

n = Number of dots in a rectangular region
length ($d_i$) = D-Number length
$d_i$ = D-Number
anchor_impact ($d_i$) = $\begin{cases} 1.4 & \text{if (anchor\_text = true)} \\ 1.0 & \text{if (anchor\_text = false)} \end{cases}$

This method of calculating the weight of regions is significant in that the number of nodes, their D-Number length (preciseness according to DDC) and their being a part of the anchor text are simultaneously taken into account. Note that substituting the outlier detection algorithm with this method will decrease the overall calculation load of the system since practically this process is performed by a simple string filtering.

If the predicted topical focus of an unvisited link complies with the specified topic of the crawler, the priority of the link should be calculated by utilizing the T-Graph. Otherwise, if the predicted topic does not match any of the topics assigned to the crawler, the link is given the least priority (0.01 by default). This is in order to force the crawler to gradually connect to and traverse other areas on the Web where there is a thick population of relevant pages.

### 5.2 Structure and Construction of T-Graph

Firstly introduced by Patel [2], the T-Graph can be described as a highly flexible tree structure in the context of graph theory. The basic idea is very much similar to that of directed graphs in that each node can have more than one parent as well as more than one child. However, their significant difference is that the nodes of T-Graph are placed in several distinct levels, as in a tree. T-Graph is also distinguished for the connection type between its nodes; each node in a specific level of the tree can directly point to the nodes of any levels below. It must be stated that in the Treasure-Crawler, the tree-like structure of the T-Graph is constructed based on the data set of interlinked Web pages that is provided by the system admin. This will help the graph to be initially constructed and made ready for the next usage. However, another possible approach is similar to that of context focused crawler [9], where the target documents are handed over to the graph and each graph automatically builds itself in a bottom-up fashion. In the initialization phase of this approach, another search engine is used to fetch the pages that point to the current document (its parents). This process continues until a desired number of levels are formed. In addition, the nodes are flexible and programmable to contain different forms of data or data structures. As shown in Fig. 5, in our proposed Treasure-Crawler system, each node of the T-Graph contains a structure of five parts that contain the following HTML attributes of a Web page:
- Immediate Sub-section Heading component (ISH).
- Heading of the Section which contains ISH (SH).
- Data Component (DC) contains the text around the link.
- Destination Information Component (DIC).

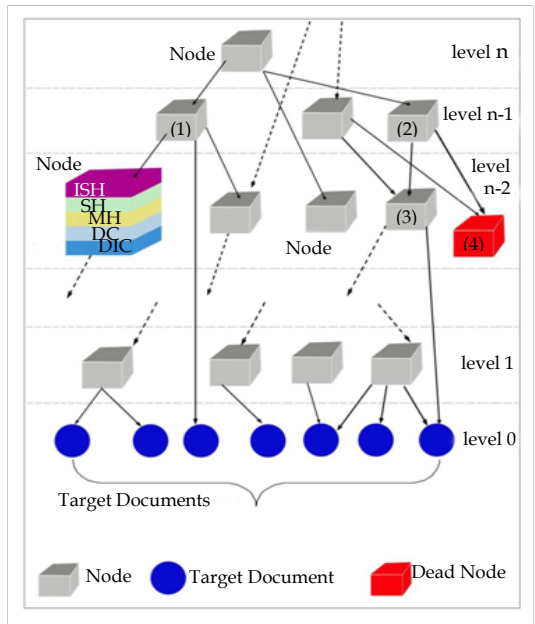

Fig. 5.T-Graph hierarchical structure

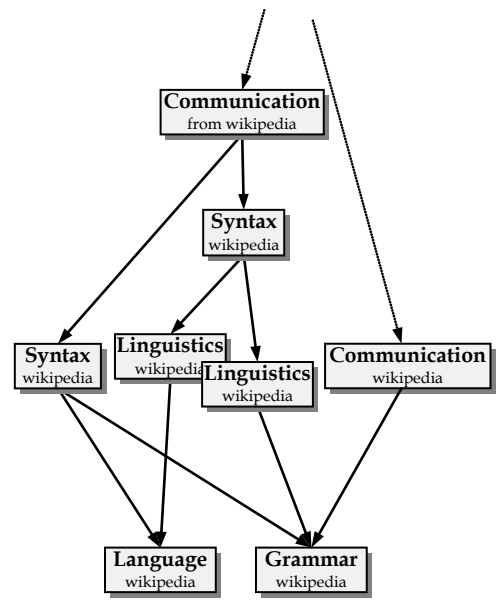

Fig. 6. T-Graph real structure from example Webpages

Practically, for each paragraph containing a hyperlink in a Web page one node is made in the T-Graph. When a paragraph contains two or more links, or a list of hyperlinks is present, one node will be made corresponding to all of them. This is the reason for a node having more than one child. Thus, for one Web page more than one node may be made in the graph. Fig. 6 shows the real T-Graph construction from example Web pages. Note that it is assumed that the links to other pages from each page are not in one common paragraph or list.

The whole concept of the T-Graph is based on the logical assumption that any unvisited link's importance is exclusively dependent on the text present within the topical boundary around it. It also assumes that if a link is present as text in the DC component of any node, it will typically point to the page containing the links, which are again represented as nodes. For information about the dead nodes and the watchdog component please refer to [2].

Therefore, the graph is constructed in two stages; in the first run the graph is made out of the pages and their interlink structure and then in the second stage the link distance of each page to reach the target level is calculated and stored in its respective component of the node. In Fig. 5 and Fig. 6 respectively, the two lines at the top do not come from a common root node because they can go up to any unknown levels or nodes in a search and retrieve cycle.

### 5.3 Calculating the Priority

As the Web page is downloaded, its unvisited hyperlinks are extracted and their topical boundaries are processed to detect the topical focus of the destination Web page in terms of D-Numbers. In order to assign a priority score to the target document, the corresponding HTML components of the Web page and those of all T-Graph nodes are compared according to the concepts and algorithms of text similarity. Each comparison gives a similarity score with the following details and all textual data is stemmed before being processed or compared.

In order to obtain specific HTML elements of the page, its source code and tag tree are analyzed by the HTML parser. Tag tree is a hierarchical representation of a Web page where the nodes are HTML tags or the text within the page. The <html> tag forms the root of the tree, and for example, anchor <a>...</a> tags may appear within paragraph <p>...</p> tags. In that case, all the text within the paragraph tags may be considered as the context of the URL within the embedded anchor tag. This context is referred to as the topical boundary. After the HTML components of the Web page are extracted, they are employed for the following processes:

The ISH data of each T-Graph node is compared with the subheading's (say U) words containing the unvisited link, to obtain a similarity measure say simISH. The subheadings' words (if any) containing the U are compared with SH data to obtain a similarity measure of simSH. The main heading of the document containing the unvisited link is compared with MH component to obtain a similarity measure of simMH. The words found within the topical boundary of the unvisited link are compared with the words present in DC component to obtain a similarity measure of simDC. The DIC component contains the information such as the least number of documents that are to be downloaded to get the target documents (link distance). The simx is the cosine similarity of two term frequency vectors [31].

Cosine similarity also normalizes the documents' length during the comparison, and is defined as the ratio of dot product to magnitude product of Vx and Vy, calculated as:

$$Sim_x(V_x, V_y) = \frac{V_x \bullet V_y}{\|V_x\|\|V_y\|} \quad (3)$$

where:
Vx is the term frequency vector, where the subscript x assumes ISH, SH, MH and DC.
Vy is the term frequency vector of the words, where y can assume either the subheading words containing the unvisited link for comparison with ISH, the subheadings containing the subheading which contains the link for comparison with SH, the main heading of the document for comparison with MH component's data and the words within topic boundary of an unvisited link for comparison with DC.
Vx•Vy is the dot product of two term frequency vectors defined as:

$$V_x \bullet V_y = \sum_{i=1}^{n} V_{x_i} \times V_{y_i} \quad (4)$$

||Vx||×||Vy|| is the product of magnitudes of vectors Vx and Vy defined as:

$$\|V_x\| \times \|V_y\| = \sqrt{\sum_{i=1}^{n}(V_{x_i})^2} \times \sqrt{\sum_{i=1}^{n}(V_{y_i})^2} \quad (5)$$

Therefore the cosine similarity is calculable as:

$$Sim_x(V_x, V_y) = \frac{V_x \bullet V_y}{\|V_x\|\|V_y\|}$$
$$= \frac{\sum_{i=1}^{n} V_{x_i} \times V_{y_i}}{\sqrt{\sum_{i=1}^{n}(V_{x_i})^2} \times \sqrt{\sum_{i=1}^{n}(V_{y_i})^2}} \quad (6)$$

Note that the cosine similarity measure of the two compared documents ranges from 0 to 1 since the term frequencies cannot be negative. Thus, the *Overall Similarity Measure (OSM)* which is a function of $sim_{ISH}$, $sim_{SH}$, $sim_{MH}$, $sim_{DC}$ for each node and the words present within the topical boundaries of an unvisited link is calculated. The *OSM* of the unvisited link is calculated as:

$$OSM = \text{average }(sim_{ISH}, sim_{SH}, sim_{MH}, sim_{DC}) \quad (7)$$

The nodes whose *OSM* is above a threshold value are considered. In our experiment, the threshold is considered as 0.05 by default which means the similarity should be at least 5% so that a node will be considered as similar. Once the matching nodes are selected, their respective DIC's data are analyzed to get the least number of documents that are to be downloaded (link distance) and depending on that value the priority of the link is calculated as:

$$Priority\ score\ = \frac{1}{min\ \{link\ distance\}} \quad (8)$$

If none of the node's *OSM* value is above the threshold, then the link is least prioritized, but still visited. The priority of such links is assigned to a fixed constant value, which is the inverse value of number of T-Graph levels incremented by 1, as studied by Diligenti et al.[9]:

$$Priority\ score\ = \frac{1}{Graph\ levels + 1} \quad (9)$$

## 6. SYSTEM TEST AND VALIDATION

Treasure-Crawler is implemented as a prototype and its performance is consequently analyzed, as well as compared to the functionality and performance of some other comparable systems, to observe its characteristics, particularly in terms of accuracy in the detection of topical focus of retrieved pages. Our prototype tests additionally produce alternative values for several parameters of the system that can be applied to its later version(s). Typically, the results show the values of precision and recall both as ~0.5. The results assert that the proposed algorithm outperforms the Context Focused Crawler in terms of accuracy with the improved number of retrieved on-topic pages by 14% for crawls with generic seed URLs and 22% for crawls with on-topic seed URLs. Details on the implementation and evaluation results of the Treasure-Crawler are elaborated in a follow-up paper titled: "Evaluation of Link and Content Based Focused Treasure-Crawler".

To target the design optimality, the framework and architecture of the proposed system has been verified against the inefficiencies of the current Web crawlers and all the functional and non-functional requirements stated in Section 3. In terms of life pilot operational test runs and gathered data, analysis and evaluation of results, the dimensions of a prototype as been implemented and tested on a fully dedicated machine with a permanent high speed Internet connection. The network, hardware and software conditions of such experimental prototype limited our evaluations to focus specifically on the Web and information retrieval common concepts; hence, we only focused on those criteria that represent the accuracy of the retrieval of relevant pages. The prototype is designed and implemented in such a way to satisfy those criteria that relate to its software engineering fundamentals such as scalability, modularity and robustness.

To assure the practicality of the system architecture, a pilot test was conducted. We built the T-Graph from an initial set of example pages and input its data into a database. One generic seed URL was supplied to the system and the parameters were initiated. This run was designated to observe the system behavior in different conditions and identify any possible bugs. The modularity and activity discretion of the modules were closely watched and verified. In some cases, we changed the initial parameters with alternate values to compare the change in outputs. The behavior of the T-Graph was also carefully monitored to test it for accuracy. Finally, this pilot test provided fruitful results and prepared the system for the major large-scale operational test runs, with the results reported in a follow-up paper. It is safe to say that we have tested, verified and validated the proposed architecture in terms of satisfying the stated requirements as well as functionality and practicality of the Treasure-Crawler system. A summary of the results follows:

• To enhance the accuracy of the topic prediction of an unvisited page, it is proved that in addition to the anchor text, taking into account some specific HTML elements of the current page is determinant.
• To decrease the processing overhead caused by the textual similarity calculations and analysis, our proposed custom method can effectively replace many statistical methods.
• To improve the quality of priority assignment to unvisited URLs, the use of T-Graph arms the system with a flexible and reliable guide map, which leads to a better determination of the links to follow.

## 7. DISCUSSION AND CONCLUSION

In this paper, we presented the architectural design of the Treasure-Crawler system. This architecture satisfies the requirements of a focused Web crawler and asserts the correctness of the system structure as well as all its modules. Based on the proposed design and the modularity of the system components, the Treasure-Crawler can easily be implemented with the waterfall model of software development process. However, the modules should first be made with respect to the object oriented or component based software development concepts. They are then weaved using specific interface to make the complete system. Throughout this process, all the necessary standards should be respected since they will affect the efficiency of the results.

In order to satisfy the objectives of this research, the architectural design of our proposed Web search system is set to comply with all the requirements of a focused Web crawler. As a result, the modules of the proposed architecture all act towards the satisfaction of two significant goals. In order to best predict the topic focus of an unvisited Web page, a custom algorithm and set of calculations are utilized based on the Dewey Decimal system.

Since the unvisited URL needs to be prioritized – to make the crawler harvest as many on-topic pages as possible – the T-Graph hierarchical structure is used as a guide to associate each unvisited Web page to a level of closeness to the sample target documents. This way, the URLs receive a score which indicates how related they are to the main topic of the crawler.

The proposed architecture is simple while thoroughly covering all its requirements. The system modules can simply be implemented based on either object oriented techniques, component based software development, or even aspect oriented techniques since they all share the support for modularity [32]. According to these facts, the outcome of the proposed architecture asserts the effectiveness of such design and therefore the system prototype is ready to be implemented.

Finally, many amendments can be applied to the proposed design. The text similarity formulas and metrics can be diversely designed and decided. HTML elements can affect the importance of a Web page differently. In conclusion, the subject field of focused Web crawlers is very interesting both theoretically and practically. To-date, there are many Web search systems designed and implemented, all targeting to overcome one major shortcoming: gathering and freshly maintaining an up-to-date index of the whole Web as efficiently and effectively as possible.

### ACKNOWLEDGEMENT

This R&D work has been conducted at the National University of Malaysia as the author's master's thesis.